\documentclass[%
 reprint,
 amsmath,amssymb,
 aps,
]{revtex4-2}

\usepackage{graphicx}
\usepackage{latexsym}
\usepackage{amssymb}
\usepackage{amsmath}
\usepackage{amsfonts}
\usepackage{upgreek}
\usepackage{bm,dsfont}% bold math
\usepackage{multirow}
\usepackage{enumitem}
\usepackage{color}
\definecolor{red}{rgb}{0.70,0.13,0.13}
\definecolor{green}{rgb}{0.13,0.55,0.13}
\definecolor{blue}{rgb}{0.25, 0.41, 0.88}
\usepackage{hyperref}
\hypersetup{
colorlinks = true,
linkcolor = red,
citecolor = green,
urlcolor = blue}
\newcommand{\bra}[1]{\langle #1|}
\newcommand{\ket}[1]{|#1 \rangle}
\newcommand{\braket}[2]{\langle #1|#2 \rangle}
\newcommand{\dd}{\mathrm{d}}
\newcommand{\ii}{\mathrm{i}}

\newcommand{\E}{\mathop{\mathbb{E}}}

\newcommand{\dsC}{\mathbb{C}}

\newcommand{\dsR}{\mathbb{R}}

\newcommand{\scE}{\mathcal{E}}

\newcommand{\scG}{\mathcal{G}}
\newcommand{\scH}{\mathcal{H}}

\newcommand{\scL}{\mathcal{L}}
\newcommand{\scM}{\mathcal{M}}
\newcommand{\scN}{\mathcal{N}}

\newcommand{\Tr}{\operatorname{Tr}}

\newcommand{\avg}{\mathop{\operatorname{avg}}}
\newcommand{\var}{\mathop{\operatorname{var}}}

\newcommand{\eq}[1]{\begin{equation}#1\end{equation}}
\newcommand{\eqs}[1]{\begin{equation}\begin{split}#1\end{split}\end{equation}}
\newcommand{\eqnref}[1]{Eq.\,\eqref{#1}}
\newcommand{\figref}[1]{Fig.\,\ref{#1}}
\newcommand{\tabref}[1]{Tab.\,\ref{#1}}
\newcommand{\secref}[1]{Sec.\,\ref{#1}}
\newcommand{\appref}[1]{Appendix\,\ref{#1}}
\newcommand{\refcite}[1]{Ref.\,\onlinecite{#1}}

\begin{document}

\title{Machine Learning Statistical Gravity from Multi-Region Entanglement Entropy}% Force line breaks with \\

\author{Jonathan Lam}
%  \altaffiliation[Also at ]{Physics Department, UCSD.}%Lines break automatically or can be forced with \\
\author{Yi-Zhuang You}%
%\email{yzyou@physics.ucsd.edu}
\affiliation{%
Department of Physics, University of California, San Diego, CA 92093, USA
%  Authors' institution and/or address\\
%  This line break forced with \textbackslash\textbackslash
}%

% \collaboration{MUSO Collaboration}%\noaffiliation

%  \homepage{http://www.Second.institution.edu/~Charlie.Author}
% \affiliation{
%  Second institution and/or address\\
%  This line break forced% with \\
% }%
% \affiliation{
%  Third institution, the second for Charlie Author
% \affiliation{%
%  Authors' institution and/or address\\
%  This line break forced with \textbackslash\textbackslash
% }%

% \collaboration{CLEO Collaboration}%\noaffiliation

\date{\today}% It is always \today, today,

\begin{abstract}
The Ryu-Takayanagi formula directly connects quantum entanglement and geometry.
Yet the assumption of static geometry lead to an exponentially small mutual information between far-separated disjoint regions, which does not hold in many systems such as free fermion conformal field theories.
In this work, we proposed a microscopic model by superimposing entanglement features of an ensemble of random tensor networks of different bond dimensions, which can be mapped to a statistical gravity model consisting of a massive scalar field on a fluctuating background geometry. 
We propose a machine-learning algorithm that recovers the underlying geometry fluctuation from multi-region entanglement entropy data by modeling the bulk geometry distribution via a generative neural network. 
To demonstrate its effectiveness, we tested the model on a free fermion system and showed mutual information can be mediated effectively by geometric fluctuation.
Remarkably, locality emerged from the learned distribution of bulk geometries, pointing to a local statistical gravity theory in the holographic bulk.
% \begin{description}
% Secondary publications and information retrieval purposes.
% You may use the \texttt{description} environment to structure your abstract;
% use the optional argument of the \verb+\item+ command to give the category of each item. 
% \end{description}
\end{abstract}

\maketitle

%\tableofcontents

\section{Introduction}

The holographic duality \cite{Brown1986CCCRASEFTG, Witten1998ASSH, Witten1998ASSTPTCGT, Gubser1998GTCFNST, Maldacena1999LLSFTS} is a duality between boundary $d$-dimensional quantum field theories and bulk $(d+1)$-dimensional gravitational theories in asymptotically anti-de Sitter (AdS) space. It provides an appealing explanation for the emergence of spacetime geometry from quantum entanglement\cite{Van-Raamsdonk2009Comments,van-Raamsdonk2010Building,Maldacena2013Cool,Jensen2013Holographic,Balasubramanian2013The-entropy,Qi2013EHMESG,Balasubramanian2014Multiboundary, Susskind2014EREPR,Balasubramanian2014Bulk,Czech2014Holographic,Cao2017Space}. The connection is manifested in the Ryu-Takayanagi (RT) formula \cite{Ryu2006Holographic,Ryu2006Aspects} $S(A)=\frac{1}{4 G_N}\min_{\gamma_A}|\gamma_A|$ that relates the entanglement entropy $S(A)$ of a boundary region $A$ to the area of the extremal surface $\gamma_A$ in the bulk that is homologous to the same region $A$. Progress has been made to reconstruct the bulk geometry from the boundary data in terms of geodesic lengths\cite{Porrati2004Boundary,Hammersley2006Extracting,Bilson2008Extracting,Cao2020Building}, extremal areas\cite{Bilson2011Extracting,Alexakis2017Determining,Bao2019Towards} or entanglement entropies\cite{You2018Machine,Roy2018Bulk}. A majority of the effort has been focused on reconstructing a classical geometry from \emph{single-region} entanglement entropies (or independent extremal surfaces). However, \emph{multi-region} entanglement entropies further encode the correlation among multiple extremal surfaces, which could reveal how the bulk geometry fluctuations around its classical background (assuming a semiclassical description of the bulk gravity). In this work, we will explore the possibility to extract information about fluctuating holographic bulk geometries from multi-region entanglement entropies of a quantum system using generative models in machine learning.

A feature of the holography entanglement entropy based on the RT formula is that the mutual information $I_{A:B} = S_A+S_B - S_{AB}$ vanishes between two disjoint boundary regions $A$ and $B$ that are far separated from each other\cite{Headrick2010Entanglement, Headrick2019Lectures}, because the minimum surface enclosing the combined region $A B$ will be a disjoint union of $\gamma_A$ and $\gamma_B$ such that the entropies simply add up as $S_{AB}=S_A+S_B$, leaving no room for mutual information. While the vanishing mutual information is a correct feature of holographic conformal field theories (CFT), it is not generally the case for many other quantum systems (e.g.~free-fermion CFT). One idea to remedy the problem is to introduce bulk matter fields to mediate the mutual information \cite{Faulkner2013Quantum,Engelhardt2015Quantum,Dong2020Effective}. Another possibility is to consider statistical fluctuations of bulk geometries such that $\gamma_A$ and $\gamma_B$ are correlated to produce the finite mutual information. The statistical gravitational fluctuation may be viewed as an effective description arising from tracing out bulk matter fields. We will further explore the second possibility of fluctuating geometry using a concrete model of random tensor network (RTN)\cite{Hayden2016Holographic,Qi2017Holographic} with fluctuating bond dimensions. The bond dimension fluctuation translates to the bulk geometry fluctuation in the context of tensor network holography\cite{Swingle2012CHSUER, Swingle2012ERH, Pastawski2015Holographic}, which is presumably governed by some statistical gravity model.

However, it is unclear what should be the appropriate bulk statistical gravity model that best reproduces the entanglement feature of a given quantum system on the boundary. To address this challenge, we propose to apply data-driven and machine-learning approaches to uncover the statistical gravity model behind the observational data of quantum many-body entanglement. What needs to be learned is the joint probability distribution of bond dimensions (or bulk geometries). Generative models\cite{Salakhutdinov2015Learning,Jimenez-Rezende2015Variational,Oord2016Pixel,Dinh2016Density,Kingma2016Improving,Oord2016WaveNet,Papamakarios2017Masked} in machine learning provide us precisely the tool to learn unknown distributions from data. In particular, we apply a deep generative model\cite{Jimenez-Rezende2015Variational,Dinh2016Density} to describe the bulk geometry fluctuation. We train the model by matching the model predictions of multi-region entanglement entropies with their actual values evaluated in the given quantum many-body state. After training, the generative model should tell us the statistical gravity model that emerges from learning. 

The approach developed in this work extends the general idea of entanglement feature learning\cite{You2018Machine}, which aims to reconstruct the bulk geometry by learning from the entanglement data on the boundary. Compare to the previous work, this study makes significant progress in including the gravitational fluctuation in the model, which will enable us to learn an emergent gravity theory rather than a static classical geometry. We will focus on (1+1)D quantum systems, and assume that the system admits an approximate semiclassical geometry description in the holographic bulk. Based on a random tensor network model with fluctuating bond dimensions, we first establish a holographic model for quantum entanglement involving a scalar matter field on a statistically fluctuating spatial geometry. Applying our approach to a free-fermion CFT state with a large central charge, we uncover a statistical gravity model governed by Weyl field fluctuations propagating on the hyperbolic background geometry. We show that the Weyl field fluctuation has the emergent bulk locality by studying its bulk correlation. We further analyze the spectrum and the leading collective modes of the emergent gravity theory. We also show that the matter field mass gets renormalized by the gravitational fluctuation as expected.

\section{Holographic Models of Entanglement}

\subsection{Random Tensor Network Model}

\begin{figure}[th]
\begin{center}
\includegraphics[width=0.95\columnwidth]{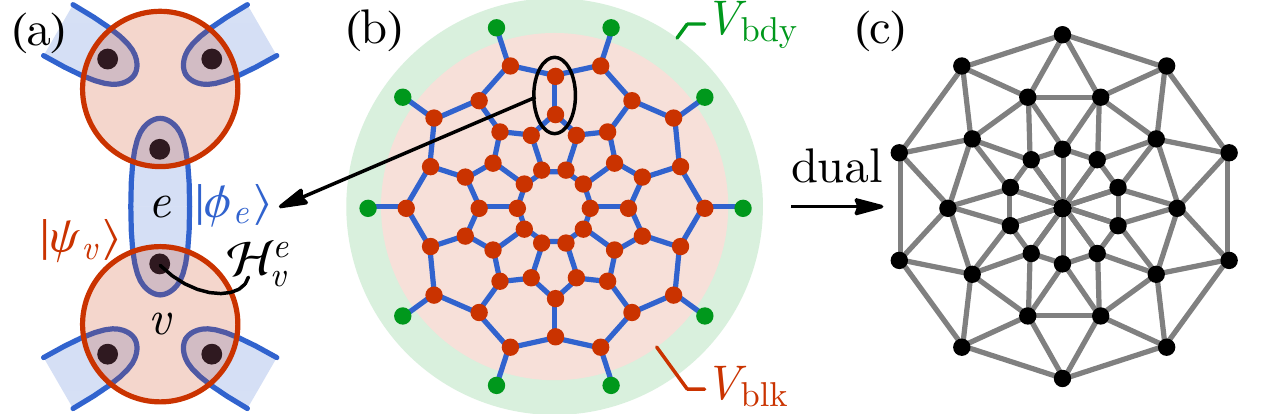}
\caption{(a) Details of the RTN near an edge. (b) The planar graph $G$ on which the RTN is defined. The vertices are classified into the bulk vertices (in red) and the boundary vertices (in green). (c) The dual graph $\tilde{G}$ of the RTN graph $G$.}
\label{fig:RTN}
\end{center}
\end{figure}

The random tensor network (RTN) model is an intuitive toy model for holographic duality, which directly connects quantum states and emergent geometries. The original proposal\cite{Hayden2016Holographic} of RTN assumes a fixed bond dimension on every link of the tensor network. It can be generalized to include bond dimension fluctuations (or more precisely, bond entanglement fluctuations)\cite{Qi2017Holographic,Vasseur2018Entanglement}. The generalized RTN model in consideration is defined as follows: (i) A planar graph $G=(V,E)$ is given to describe the background network geometry, where $V$ denotes the vertex set and $E$ denotes the edge set. $V=V_\text{blk}\cup V_\text{bdy}$ is divided into two subsets: the bulk $V_\text{blk}$ and the boundary $V_\text{bdy}$ sets, see \figref{fig:RTN}(b). (ii) A local Hilbert space $\scH_{v}^{e}$ is associated with each pair $(v,e)$ of vertex $v\in V$ and its adjacent edge $e\in E$ (for $e$ not adjacent to $v$, the associated Hilbert space is considered trivial $\scH_{v}^{e}\cong \dsC$), see \figref{fig:RTN}(a). (iii) A random pure state $\ket{\psi_v}\in\scH_{v}\equiv\bigotimes_{e\in\dd v}\scH_{v}^{e}$ is defined on every bulk vertex $v\in V_\text{blk}$. (iv) A random entangled state $\ket{\phi_e}\in\scH^{e}\equiv\bigotimes_{v\in\partial e}\scH_{v}^{e}$ is defined across every edge $e\in E$. (v) RTN defines an ensemble $\scE_\text{RTN}=\{\ket{\Psi}\}$ of pure states in the boundary Hilbert space $\scH_\text{bdy}\equiv\bigotimes_{v\in V_\text{bdy}}\scH_{v}$ by taking a (partial) projection in the bulk Hilbert space $\scH_\text{blk}\equiv\bigotimes_{v\in V_\text{blk}}\scH_{v}$ as
\eq{\ket{\Psi}=\braket{\psi}{\phi}:
\ket{\psi}=\bigotimes_{v\in V_\text{blk}}\ket{\psi_v},\ket{\phi}=\bigotimes_{e\in E}\ket{\phi_e}.}
The probability measure of $\ket{\Psi}$ in the RTN ensemble $\scE_\text{RTN}$ is given by $P(\ket{\Psi})=P(\ket{\psi})P(\ket{\phi})$. The vertex state distribution $P(\ket{\psi})=\prod_{v\in V_\text{blk}}P(\ket{\psi_v})$ is assumed to be factorized, and on each vertex, the distribution $P(\ket{\psi_v})$ is taken to be the Haar measure (i.e.~uniform random states in $\scH_{v}$). The edge (link) state distribution $P(\ket{\phi})$ is generally a nontrivial joint distribution depending on all $\ket{\phi_e}$ on all edges, which allows the quantum entanglement across different edges to fluctuate collectively.

For any operator $O^{(k)}$ defined in $k$ copies of the boundary Hilbert space $\scH_\text{bdy}^{\otimes k}$, its expectation value in the product state $\ket{\Psi}^{\otimes k}$ is defined to be
\eq{\langle O^{(k)}\rangle =\E_{\ket{\Psi}\in\scE_\text{RTN}}\frac{\Tr\big((\ket{\Psi}\bra{\Psi})^{\otimes k}O^{(k)}\big)}{\braket{\Psi}{\Psi}^k}.}
We assume that the correlation between denominator and numerator is not important (which is generally valid in the semiclassical regime when fluctuations are weak), so that we can approximate the ensemble average of the ratio by the ratio of separate averages,
\eq{\langle O^{(k)}\rangle \simeq \frac{1}{\scN_{k}}\E_{\ket{\Psi}\in\scE_\text{RTN}}\Tr\big((\ket{\Psi}\bra{\Psi})^{\otimes k}O^{(k)}\big),}
where $\scN_k=\E_{\ket{\Psi}\in\scE_\text{RTN}}\braket{\Psi}{\Psi}^k$ is the $k$th moment of the state norm squared. For example, the 2nd R\'enyi entanglement entropy $S_A$ (or more precisely, the purity $e^{-S_A}$) of RTN states in a boundary region $A$ can be calculated by taking $k=2$ and $O^{(k)}=X_A$ (the swap operator supported in region $A$),
\eq{\label{eq:EF1}e^{-S_A}\propto \E_{\ket{\Psi}\in\scE_\text{RTN}}\Tr\big((\ket{\Psi}\bra{\Psi})^{\otimes 2}X_A\big).}
We will suppress the R\'enyi index throughout this work, and use $S_A$ to denote the 2nd R\'enyi entropy. The RTN model provides an effective description of entanglement entropies of typical quantum states on the holographic boundary, given the background geometry $G$ together with fluctuations of states $\ket{\psi_v}, \ket{\phi_e}$ in the holographic bulk.

It worth mention that in modeling the 2nd R\'enyi entanglement entropy by \eqnref{eq:EF1}, the average over the RTN ensemble $\scE_\text{RTN}$ is taken neither on the state vector level (i.e.~not a pure state superposition $\E_{\ket{\Psi}\in\scE_\text{RTN}}\ket{\Psi}$), nor on the density matrix level (i.e.~not a mixed state superposition $\E_{\ket{\Psi}\in\scE_\text{RTN}}\ket{\Psi}\bra{\Psi}$), but on the double density matrix level (as $\E_{\ket{\Psi}\in\scE_\text{RTN}}(\ket{\Psi}\bra{\Psi})^{\otimes 2}$). The same average strategy commonly appeared in random tensor network/quantum circuit literatures\cite{Hayden2016Holographic,Qi2017Holographic,You2018Machine,Bao2019Holographic,Kuo2019Markovian,Fan2021Self-organized}. Such average may not have direct physical realization, nevertheless it defines a RTN model for entanglement entropy which can produce (i) positive mutual information $I_{A:B}$ that does not vanish between distant regions and (ii) possibly negative tripartite information $I_{A:B:C}$ (see \appref{app:perturbative} for a perturbative proof). These features indicate that the generalized RTN model is expressive enough to describe quantum chaotic states with information scrambling\cite{Hosur2016Chaos,Seshadri2018Tripartite} and to capture mutual information between distant entanglement regions, which goes beyond holographic CFT states.

\subsection{Ising and Dual Ising Models}

Evaluating the ensemble average in \eqnref{eq:EF1} following the approach developed in \refcite{Hayden2016Holographic}, the RTN purity $e^{-S_A}$ can be map to the partition function of an Ising model on the graph $G$ with fluctuating coupling constants
\eq{\label{eq:EF2}
e^{-S_A}=\sum_{[\sigma,J]}P[\sigma|J]P[J]\,\delta[\sigma_\text{bdy}\Leftrightarrow A],}
with $P[\sigma |J]$ given by
\eqs{P[\sigma |J]&=\frac{e^{-E[\sigma|J] }}{Z[J]},\\
E[\sigma|J]&=-\sum_{e\in E} \bigg(\frac{J_e}{2} \prod_{v\in\partial e}\sigma_v\bigg),\\
Z[J]&=\sum_{[\sigma]}e^{-E[\sigma|J]}\delta[\sigma_\text{bdy}\Leftrightarrow\emptyset].}
and $P[J]$ given by
\eq{P[J]=\int_{\ket{\phi}}\prod_{e\in E}\delta\big(J_e-S(\ket{\phi_e})\big)P(\ket{\phi}).}
Here $\sigma_v=\pm1$ is the Ising variable defined on every vertex $v\in V$, $J_e\geq 0$ is the ferromagnetic coupling strength on every edge $e\in E$. $J_e$ is determined by $S(\ket{\phi_e})$, the 2nd R\'enyi entropy of the state $\ket{\phi_e}$ (entangled between the Hilbert spaces $\scH_{v_{+} e}$ and $\scH_{v_{-} e}$ where $v_{\pm}$ are the two vertices on the boundary of $e$). $J_e$ characterizes how much the tensors are entangled with each other across the edge $e$ in the tensor network. It corresponds to the notion of bond dimension when $\ket{\phi_e}$ is maximally entangled. The distribution $P[J]$ describes the how the effective bond dimension (bond entanglement) fluctuates in the RTN ensemble. Finally, the partition function is subject to the boundary condition that is set by the boundary region $A$ of $S_A$,
\eq{\forall v\in V_\text{bdy}:\sigma_v=\left\{
\begin{array}{cc}
+1 & v\notin A,\\
-1 & v \in A,
\end{array}\right.}
which is denoted as $\delta[\sigma_\text{bdy}\Leftrightarrow A]$ in \eqnref{eq:EF2}. The partition function $Z[J]$ properly normalizes the Boltzmann weight of the Ising model, such that $S_A=0$ when the entanglement region $A=\emptyset$ is empty.

Given that $G$ is a planar graph\footnote{The model can be more expressive if the planar graph assumption is lifted, however it will be challenging to make connection to the RTN model on non-planar graphs.}, we can use the Kramers-Wannier duality to rewrite the Ising model \eqnref{eq:EF2} on the dual lattice $\tilde{G}=(\tilde{V},\tilde{E})$, as shown in \figref{fig:RTN}(c), where $\tilde{V}$ corresponds to the set of faces in $G$ and $\tilde{E}\cong E$. The dual Ising model takes the similar form
\eqs{\label{eq:EF3}
e^{-S_A}=\sum_{[\tilde{\sigma},\tilde{J}]}\bigg(\prod_{\tilde{v}\in\partial A}\tilde{\sigma}_{\tilde{v}}\bigg) P[\tilde{\sigma}|\tilde{J}]P[\tilde{J}],}
with $P[\tilde{\sigma}|\tilde{J}]$ given by
\eqs{P[\tilde{\sigma}|\tilde{J}]&=\frac{e^{-E[\tilde{\sigma}|\tilde{J}]}}{Z[\tilde{J}]},\\
E[\tilde{\sigma}|\tilde{J}]&=-\sum_{\tilde{e}\in \tilde{E}} \bigg(\frac{\tilde{J}_{\tilde{e}}}{2} \prod_{\tilde{v}\in\partial \tilde{e}}\tilde{\sigma}_{\tilde{v}}\bigg),\\
Z[\tilde{J}]&=\sum_{[\tilde{\sigma}]}e^{-E[\tilde{\sigma}|\tilde{J}]},}
and $P[\tilde{J}]$ related to $P[J]$ by
\eq{P[\tilde{J}]=\bigg(\prod_{e}\frac{\partial J_e}{\partial \tilde{J}_{\tilde{e}}}\bigg) P[J].}
Here $\tilde{\sigma}_{\tilde{v}}=\pm 1$ is the dual Ising variable and $\tilde{J}_{\tilde{e}}=-\ln \tanh(J_e/2)$ is the dual coupling. The boundary condition in the original Ising model translates to the insertion of the dual Ising variable at every boundary point of entanglement region $A$ (i.e.~at every entanglement cut). The partition function $Z[\tilde{J}]$ on the denominator ensures $S_A=0$ when the entanglement region $A=\emptyset$ is empty, i.e.~when there is no insertion of dual Ising variables. Both $P[\tilde{\sigma}|\tilde{J}]$ and $P[\tilde{J}]$ are normalized probability distributions, which defines the joint distribution $P[\tilde{\sigma},\tilde{J}]=P[\tilde{\sigma}|\tilde{J}]P[\tilde{J}]$ for dual Ising variables and their couplings. Therefore the purity of the RTN state can be interpreted as the boundary correlation of dual Ising variables $e^{-S_A}=\langle\prod_{\tilde{v}\in\partial A}\tilde{\sigma}_{\tilde{v}}\rangle$ in an Ising model with fluctuating couplings. 

\begin{figure}[h]
\begin{center}
\includegraphics[width=0.62\columnwidth]{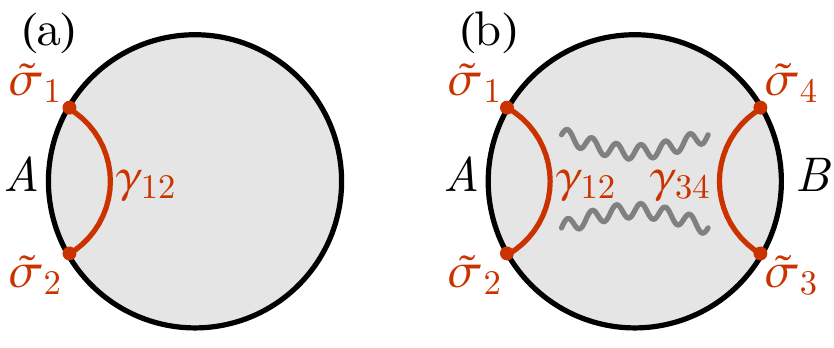}
\caption{(a) Two-point correlation and (b) four-point correlation of dual Ising spins.}
\label{fig:correlation}
\end{center}
\end{figure}

The RT formula can be recovered in the classical limit when the RTN bond dimensions are large and fixed, which corresponds to the deep ferromagnetic phase of the original Ising model ($J_e\gg 1$) or equivalently the deep paramagnetic phase of the dual Ising model ($\tilde{J}_{\tilde{e}}\ll 1$). In such limit, the dual Ising correlation decays exponentially with the geodesic distance $\langle\tilde{\sigma}_1\tilde{\sigma}_2\rangle\propto e^{-|\gamma_{12}|/\xi}$, as illustrated in \figref{fig:correlation}(a), which reproduces the RT formula $S_A=|\gamma_{12}|/\xi$ with some appropriate choice of the correlation length $\xi$. Multi-region entanglement entropies will correspond to higher-point correlations functions, such as $e^{-S_{AB}}\sim\langle\tilde{\sigma}_1\tilde{\sigma}_2\tilde{\sigma}_3\tilde{\sigma}_4\rangle\sim e^{-|\gamma_{12}|/\xi}e^{-|\gamma_{34}|/\xi}$ in \figref{fig:correlation}(b). Allowing the dual Ising coupling $\tilde{J}$ to fluctuate collectively will introduce perturbations to the geodesic distance $|\gamma_{12}|\to |\gamma_{12}|+\delta|\gamma_{12}|$ in a correlated manner, such that
\eqs{e^{-S_{AB}}&\sim \E\big(e^{-(|\gamma_{12}|+\delta|\gamma_{12}|)/\xi}e^{-(|\gamma_{34}|+\delta|\gamma_{34}|)/\xi}\big)\\
&\sim e^{-|\gamma_{12}|/\xi} e^{-|\gamma_{34}|/\xi} \;e^{\frac{1}{2\xi^2}\E\delta|\gamma_{12}|\delta|\gamma_{34}|}\\
&\sim e^{-S_A}e^{-S_{B}}e^{I_{A:B}}.}
Thus the correlated geometric fluctuation provides an effective mechanism to generate the mutual information between far-separated regions $A$ and $B$ (beyond the classical RT formula). Therefore we anticipate the fluctuating RTN model to be a more expressive holographic model for entanglement entropies. However, it is not clear how the dual Ising coupling $\tilde{J}$ (or the effective bond dimension $J$) should fluctuate precisely in order to quantitatively reproduce all multi-region entanglement entropies of a given quantum many-body state. The remaining task is learn the distribution $P[\tilde{J}]$ (or other equivalent distributions) from data.

\subsection{Effective Statistical Gravity Model} 

Suppose the fluctuation of $\tilde{J}$ is small around its static background configuration, such that there is a meaningful notion of background geometry in the bulk. The dual Ising model can be described by an effective field theory in the continuum limit
\eq{S[\phi|g]=\frac{1}{2}\int\dd^2 x\sqrt{g}(g^{ij}\partial_i \phi\partial_j\phi+m^2\phi^2),}
where the dual Ising variable $\tilde{\sigma}_{\tilde{v}}$ is coarse-grained to a massive real scalar field $\phi(x)$, as the Ising model universally flows to this massive Gaussian fixed point in the paramagnetic phase. The theory is defined in the holographic space (without time dimension). The fluctuating Ising coupling $\tilde{J}$ can be translated to a fluctuating bulk metric tensor $g$ around a reference background geometry $\bar{g}$ \footnote{Another interpretation is to translate the fluctuating Ising coupling to the fluctuating mass term, as the scalar field mass is the relevant perturbation that drives the order-disorder transition, which plays the same role as the Ising coupling. This alternative view turns out to be equivalent to the fluctuating metric interpretation in two-dimension, as we will see soon.}, since a stronger local coupling creates a larger local correlation, which effectively reduces the local distance measure $\dd s^2=g_{ij}\dd x^i \dd x^j$ between the correlated Ising variables. Therefore the purity of RTN state \eqnref{eq:EF3} can be effectively described by a statistical gravity model
\eqs{\label{eq:EF4}
e^{-S_A}&=\int_{[\phi,g]} \bigg(\prod_{x\in\partial A}\phi(x)\bigg) P[\phi|g]P[g],\\
P[\phi|g]&=\frac{e^{-S[\phi|g]}}{Z[g]}, Z[g]=\int_{[\phi]}e^{-S[\phi|g]},}
where the gravity is ``quenched'' in the sense that the metric configuration is generated with a probability distribution $P[g]$ independent of the scalar field $\phi$ configuration.

In two-dimensional space, the metric tensor has three independent components. However two of them can be removed by gauge transformation $g_{ij}\to g_{ij}+\nabla_i\xi_j+\nabla_j\xi_i$. We can choose the conformal gauge where the metric tensor $g_{ij}(x)$ is parametrized by a Weyl field $\omega(x)$ that rescales a fixed background $\bar{g}_{ij}(x)$
\eq{g_{ij}(x)=e^{2\omega(x)}\bar{g}_{ij}(x),}
such that each Weyl field configuration represents a physically distinct geometry. As a result, the integration $\int_{[g]}P[g]$ can be replaced by $\int_{[\omega]}P[\omega]$ in \eqnref{eq:EF4}. The unknown joint distribution $P[\omega]$ will be what we aim to learn from the entanglement entropy data.

To numerically evaluate the multi-point scalar field correlation, we can place the bulk field theory back on a lattice, say on the dual graph $\tilde{G}=(\tilde{V},\tilde{E})$. Using Regge calculus\cite{Regge1961General} to discretize the action, 
\eq{\label{eq:S_phi}
S[\phi|\omega] =\sum_{\langle xy\rangle\in\tilde{E}}\frac{A_{xy}}{2}\Big(\frac{\phi_{x}-\phi_{y}}{\ell_{xy}}\Big)^2+\sum_{x\in\tilde{V}}\frac{m^2A_{x}}{2}e^{2\omega_{x}}\phi_{x}^2,}
where $\ell_{xy}$ can be interpreted as the geodesic distance between two vertices $x$ and $y$ on the background geometry. $A_{x}$ and $A_{xy}$ are the areas associated to the vertex $x$ and the edge $\langle xy\rangle$ respectively. $\ell_{xy}, A_{x},A_{xy}$ are all fixed according to the choice of background metric, which will be specified later. The statistical variables in the model are the scalar field $\phi_x$ and the Weyl field $\omega_x$ in the holographic bulk. The model predicts the entanglement entropy on the holographic boundary by
\eqs{\label{eq:EF5}
e^{-S_A}&=\int_{[\phi,\omega]} \bigg(\prod_{x\in\partial A}\phi_{x}\bigg) P[\phi|\omega]P[\omega],\\
P[\phi|\omega]&=\frac{e^{-S[\phi|\omega]}}{Z[\omega]}, Z[\omega] = \int_{[\phi]}e^{-S[\phi|\omega]}}
which is the underlying lattice model that will be used in the machine learning algorithm. The unknown distribution $P[\omega]$ will be parameterized by a generative model. By matching the model prediction with the actual data of entanglement entropies calculated from a quantum state, the algorithm can reconstruct the distribution $P[\omega]$ and infer the statistical gravity model behind the entanglement structure.

\section{Machine Learning Method}

\subsection{Generative Modeling}

Generative modeling is about learning probability distributions\cite{Goodfellow2016DL}. We will apply the simplest latent-variable generative model \cite{Goodfellow2014Generative} in this work. The basic idea is to start with a easy-to-sample prior distribution, such as a Gaussian distribution. Draw a random vector $z\in \dsR^n$ (as a collection of latent variables) from the prior distribution $P(z)$. Then transform the latent variables $z$ by a deep neural network $G_\vartheta$ (parametrized by some variational parameter $\vartheta$) to the designated random variable $\omega$, i.e.~$z\to\omega = G_\vartheta(z)$. The mapping $G_\vartheta$ is called the generator, which defines the distribution of generated samples
\eq{\label{eq:Ptheta}P_\vartheta[\omega]=\int\dd z\;\delta(\omega-G_\vartheta(z))P(z).}
A large batch of $\omega$ can be sampled efficiently in parallel, when hardware accelerators (e.g GPU or TPU) are available. If the neural network $G_\vartheta$ is expressive enough, \eqnref{eq:Ptheta} will provide a sufficiently expressive probability model $P_\vartheta[\omega]$ for the Weyl field $\omega$ configuration.

The distribution $P_\vartheta[\omega]$ defines the model prediction of the purity based on \eqnref{eq:EF5}
\eq{\label{eq:EF6}
e^{-S_A|_\vartheta}=\E_{[\omega]\sim P_\vartheta}\langle\textstyle\prod_{\partial A}\phi\rangle_\omega.}
We will use $S_A|_\vartheta$ to denote the R\'enyi entropy predicted by the machine learning model as it depends on the model parameters $\vartheta$. In \eqnref{eq:EF6}, we introduced the short-hand notation
\eq{\langle{\textstyle\prod_{\partial A}}\phi\rangle_\omega \equiv \int_{[\phi]} \bigg(\prod_{x\in\partial A}\phi_{x}\bigg) P[\phi|\omega]}
to denote the scalar field correlation on a background Weyl field configuration. The conditional distribution $P[\phi|\omega]$ is defined in \eqnref{eq:EF5} with $S[\phi|\omega]$ given in \eqnref{eq:S_phi}. The scalar field correlation $\langle{\textstyle\prod_{\partial A}}\phi\rangle_\omega$ can be efficiently evaluated when $S[\phi|\omega]$ is a Gaussian action (which is the case here).

\begin{figure}[bh]
\begin{center}
\includegraphics[width=0.72\columnwidth]{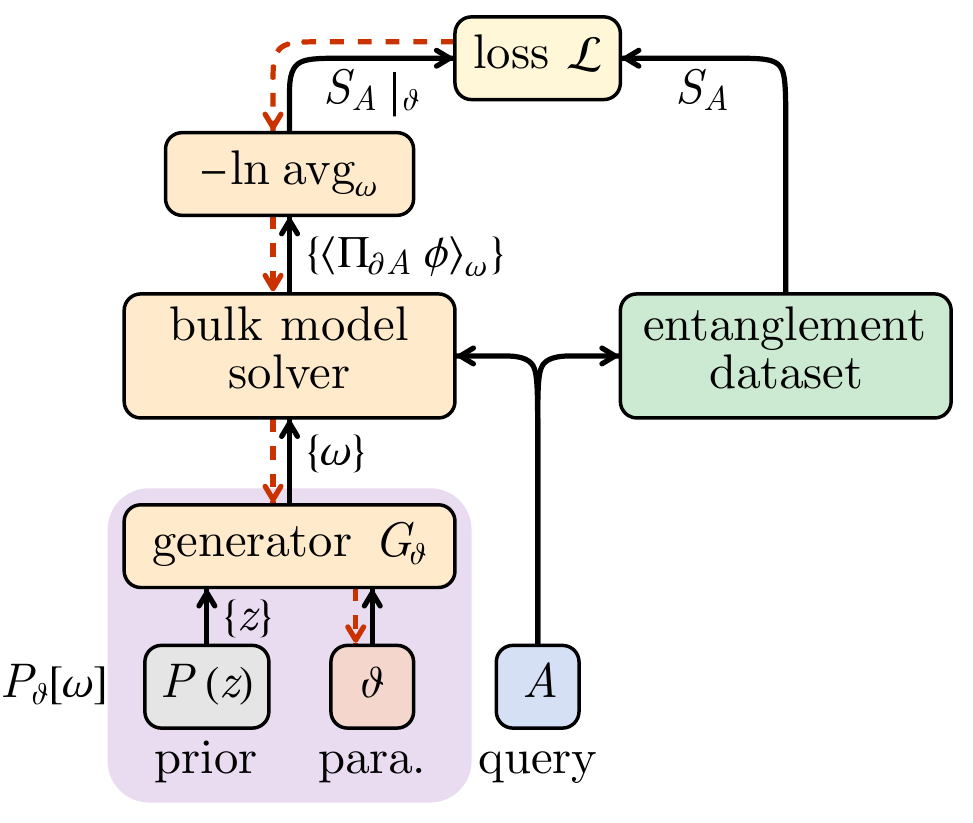}
\caption{Flow diagram of the machine learning algorithm. Black arrows denotes the forward evaluation of the loss function. Red dashed arrows denotes the gradient back propagation to train the parameter.}
\label{fig:flow}
\end{center}
\end{figure}

The task is to learn the optimal Weyl field distribution $P_\vartheta[\omega]$ that gives the best prediction of the purity data based on \eqnref{eq:EF6}. The dataset will contain the purity $\{e^{-S_A}\}$ of a quantum state in different regions $A$. The distribution $P_\vartheta[\omega]$ can be learned by optimizing model parameters $\vartheta$ to minimize the following loss function (to be explained later)
\eq{\label{eq:loss}\scL_\vartheta = \avg_{A}\big(1-e^{S_A|_\vartheta-S_A}\big)^2.} 
As illustrated in \figref{fig:flow}, the training initiates from randomly choosing a batch of entanglement regions $A$. On one hand, we query the dataset to get the ground truth of $S_A$. On the other hand, a collection of Weyl field configurations are sampled from the generative model, based on which the model prediction $S_A|_\vartheta$ is estimated. Then the loss function is calculated by comparing $S_A|_\vartheta$ with $S_A$, and the gradient signal propagates back to train the parameters via gradient descent $\vartheta\to\vartheta- r \partial_\vartheta \scL_\vartheta$. After some iterations, the parameters are expected to converge. In the following, we will explain different modules in \figref{fig:flow} in detail.
 
\subsection{Entanglement Dataset}

While efficient experimental approaches\cite{Brydges2019Probing,Huang2020Predicting} have been developed to estimate R\'enyi entropies from randomized measurements, which enables the acquisition of a large amount of entanglement data to drive the entanglement feature learning, preparing an entanglement dataset by numerically computing entanglement entropies from a given quantum many-body state remains difficult in general. As a proof of concept, we choose to use the ground state of a free fermion system for demonstration, on which entanglement entropies can be efficiently calculated. 

Consider $N$ copies of the (1+1)D massless Majorana fermion chain, described by the Hamiltonian 
\eq{
H = \sum_{a=1}^{N} \sum_j \ii \chi_{j,a} \chi_{j+1,a},}
where $\{ \chi_{i,a}, \chi_{j,b} \} = \delta_{ij} \delta_{ab} $. Let $\ket{\Psi}$ be the ground state of $H$. The 2nd R\'enyi entropy can be efficiently computed from the fermion correlation function,
\eq{\label{eq:S_free_fermion}S_A= -\frac{1}{2}\Tr \ln(C_A^2+(1-C_A)^2),}
where $C_{A,ij}=\bra{\Psi}\chi_{i,a}\chi_{j,a}\ket{\Psi}$ (for $i,j\in A$) is the two-point correlation function (matrix) of Majorana fermions restricted inside the entanglement region $A$. The quantum system is critical and is described by the free-fermion CFT at low energy. 
%For a single region $A$, the asymptotical behavior of the $n$th-Renyi entropy grows logarithmically with the region size $|A|$ as \eq{S_A^{(n)} = \frac{c}{6}\Big(1+\frac{1}{n} \Big) \ln \Big( \frac{|A|}{\delta} \Big),} where $\delta$ is a non-universal UV-cutoff constant and $c=N/2$ is the total central charge.

To construct the dataset, we will take the Majorana fermion chain of 32 sites, and randomly sample a large collection of single-region, two-region, and three-region subsets. We then compute the entanglement entropy using \eqnref{eq:S_free_fermion} for every region and record the results in the entanglement dataset. 

\subsection{Bulk Model Solver}

The bulk model solver is expected to calculate the scalar field correlation given the Weyl field background $\omega$ and the entanglement region $A$ that specifies the scalar field inserting position on the boundary. We will use the lattice model specified by the action in \eqnref{eq:S_phi}, which describe a free scalar field $\phi$. The action can be written as the bilinear form
\eq{\label{eq:S_phi|omega}S[\phi|\omega]=\frac{1}{2}\sum_{x,y\in \tilde{V}}\phi_{x}K^{(\phi)}_{xy}[\omega]\phi_{y}}
where $x,y$ label the vertices on the dual graph $\tilde{G}=(\tilde{V},\tilde{E})$ on which the holographic model is defined. The kernel matrix takes the form of $K^{(\phi)}[\omega]=\nabla^2+M[\omega]$, with \eq{M_{xy}[\omega]=m^2A_x e^{2\omega_x}\delta_{xy}} being the mass term, and \eq{\nabla^2_{xy}=\sum_{\langle x'y'\rangle\in\tilde{E}}\frac{A_{x'y'}}{\ell_{x'y'}^2}(\delta_{xx'}-\delta_{xy'})(\delta_{yx'}-\delta_{yy'})} being the discrete Laplace operator on the dual graph. The length $\ell_{xy}$ and area $A_x,A_{xy}$ constants are fixed and are set by the background geometry, as to be specified soon. The two-point correlation is given by the inverse of the kernel matrix,
\eq{\langle \phi_x\phi_y\rangle_\omega=\frac{1}{\beta}((K^{(\phi)}[\omega])^{-1})_{xy},}
where a trainable constant $\beta$ is introduced to take care of the field renormalization. Higher-point correlations follow from Wick's theorem. For example,
\eqs{\langle\phi_x\phi_y\phi_z\phi_w\rangle_\omega&=
\langle\phi_x\phi_y\rangle_\omega\langle\phi_z\phi_w\rangle_\omega\\
&+\langle\phi_x\phi_z\rangle_\omega\langle\phi_y\phi_w\rangle_\omega\\
&+\langle\phi_x\phi_w\rangle_\omega\langle\phi_y\phi_z\rangle_\omega.}
We will treat $\beta$ and $m^2$ as trainable parameters, which will be optimized (together with other model parameters for $P[\omega]$) to fit the entanglement data.

Since we intend to apply our approach to entanglement data collected from CFTs, following the idea of AdS/CFT correspondence, it is natural to choose the two-dimension hyperbolic geometry (the spatial slice of AdS$_3$) as the background geometry. We use the following background metric
\eq{\dd s^2=\dd\rho^2+\sinh^2\rho\;\dd \theta^2,}
where $0\leq \theta < 2\pi$ and $\rho\leq \rho_\text{bdy}$ (the UV cutoff scale is set by $\rho_\text{bdy}$, which is another parameter to learn).  
The geodesic distance any two points on the boundary $\rho=\rho_\text{bdy}$ separated by $\theta$ is given by 
\eqs{\label{eqn:geodesic}
|\gamma|(\theta) &= \text{arccosh} (1 +2 \sinh^2\rho_\text{bdy} \sin^2 (\theta/2) ) \nonumber \\ 
&\xrightarrow{e^{\rho_\text{bdy}} \gg 1} 2 \ln (\sin (\theta/2) +\rho_\text{bdy}).}

\begin{figure}[t]
\begin{center}
\includegraphics[width=0.52\columnwidth]{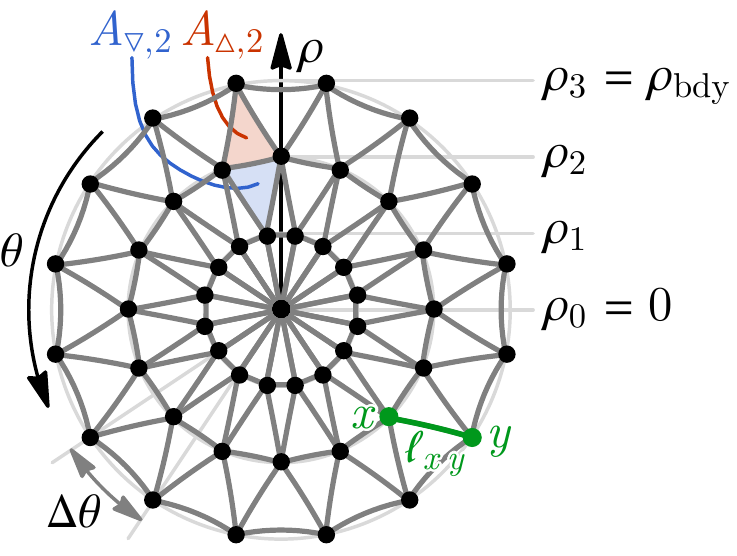}
\caption{Triangular lattice discretization of hyperbolic space in $(\rho,\theta)$ coordinate. The lattice is divided into different layers along the the radius direction. The $i$th layer corresponds to the radius $\rho_i$.}
\label{fig:lattice}
\end{center}
\end{figure}

Without loss of generality, we chose to discretize space using a triangular lattice with periodic boundary condition along the $\theta$-direction. All vertices in the same layer are of the same $\rho$-coordinate and their $\theta$-coordinates are uniformly spaced, see \figref{fig:lattice}. The geodesic distance $\ell_{xy}$ between two vertices $x$ and $y$ is given by
\eqs{\cosh\ell_{xy}&=\cosh\rho_x\cosh\rho_y\\
&-\sinh\rho_x\sinh\rho_y\cos(\theta_x-\theta_y).}
The area of an elementary triangle in the $i$th layer reads
\eqs{\tan\frac{A_{\bigtriangleup,i}}{4}&=\tanh\Big(\frac{\rho_{i+1}-\rho_{i}}{2}\Big)\tanh\Big(\frac{b_i}{4}\Big),\\
\tan\frac{A_{\bigtriangledown,i}}{4}&=\tanh\Big(\frac{\rho_{i}-\rho_{i-1}}{2}\Big)\tanh\Big(\frac{b_i}{4}\Big),\\
\cosh b_i&=\cosh^2\rho_i -\sinh^2\rho_i\cos\Delta\theta,} 
which defines the vertex and edge areas in a barycentric scheme. Specifically, the vertex area $A_x$ is given by
\eqs{A_x=\tfrac{1}{3}(2A_{\bigtriangleup,i}+2A_{\bigtriangledown,i}+A_{\bigtriangleup,i-1}+A_{\bigtriangledown,i+1}),}
for $\rho_x=\rho_i$. The edge area $A_{xy}$ is given by
\eq{A_{xy}=\left\{
\begin{array}{ll}
\tfrac{1}{3}(A_{\bigtriangleup,i}+A_{\bigtriangledown,i}) & \rho_x=\rho_y=\rho_i;\\
\tfrac{1}{3}(A_{\bigtriangleup,i}+A_{\bigtriangledown,i+1}) & \rho_x=\rho_i, \rho_y=\rho_{i+1};\\
\tfrac{1}{3}(A_{\bigtriangleup,i-1}+A_{\bigtriangledown,i}) & \rho_x=\rho_i, \rho_y=\rho_{i-1}.\\
\end{array}\right.}
These equations defines $\ell_{xy}$, $A_{x}$ and $A_{xy}$ used in the lattice model \eqnref{eq:S_phi}, which all rely on the values of $\rho_i$ for different layers. The discretization scheme in the radial dimension is specified by how $\rho_i$ is spaced from $0$ to $\rho_\text{bdy}$. A bad choice of the discretization scheme may cause some triangle elements to have high aspect ratios, reducing the quality of the triangulation in approximating the continuous background geometry. We will take a data-driven approach to learn the optimal discretization scheme by treating $\{\rho_i\}$ as trainable parameters. 

To summarize, the bulk model solver contains the following parameters: the normalization $\beta$ and the squared mass $m^2$ associated with the scalar field dynamics, and the radial coordinates $\{\rho_i\}$ associated with the discretization of background geometry. These parameters will be trained together with other neural network parameters (see \secref{sec:NN}) to optimize the model prediction of the entanglement entropy data.

\subsection{Neural Network Design}\label{sec:NN}

The central goal is to learn the Weyl field distribution $P[\omega]$ using a latent-variable generative model $P_\vartheta[\omega]$, recall \eqnref{eq:Ptheta}. The key component of the generative model is a generator $G_\vartheta$ that maps the latent variable $z$ to a Weyl field configuration $\omega=G_\vartheta(z)$. The generator is realized as a deep neural network consists of consecutive layers of simpler maps 
\eq{G_\vartheta(z)=g_N \circ \cdots g_2 \circ g_1 (z),}
where each layer $g_n(z)$ is an affine transformation followed by some non-linearity such as ReLU \cite{Agarap2018Deep}. The weight and bias parameters are introduced to parametrize the affine transformations, which constitute part of the training parameters $\vartheta$.

It is both practical and theoretically motivated to enforce the neural network's architecture such that the learned distribution $P_\vartheta[\omega]$ will respects certain symmetries, i.e.~to construct an equivariant neural network \cite{Cohen2016Group}. Let $Q$ be a symmetry transformation that we wish to impose. The sufficient condition for the generated distribution to be symmetric (i.e. $P_\vartheta[Q\omega]=P_\vartheta[\omega]$) is to require (i) $Qg_n(z) = g_n(Qz)$ and (ii) $P(Qz) = P(z)$. The symmetries in consideration are
\begin{enumerate}
  \item $\omega_{(\rho, \theta)} \to \omega_{(\rho, \theta+a)} $ [translation],
  \item  $\omega_{(\rho, \theta)} \to \omega_{(\rho, -\theta)} $ [reflection].
\end{enumerate}
The translation symmetry can be imposed by parameter sharing between relation-related weights and biases, making the affine transformation in each layer effectively a convolution along the translation direction. The reflection symmetry can be imposed by using a reflection symmetric convolution kernel. The prior distribution $P(z)=\prod_xP(z_x)$ automatically satisfies the symmetry condition as it factorizes to identical independent Gaussian distributions on every site.

We would like to emphasize that although each layer looks like a convolutional layer under the symmetry constraint, we do not restrict the convolution kernel to be local (the kernel size extends to the whole lattice), because we do not want to impose locality by hand. As we will see, a sense of locality could emerge in the neural network as the holographic model gets trained, which corresponds to the emergent locality in the bulk gravity theory. 

\subsection{Loss Function Design} 

The loss function is designed to evaluate the average difference between the purity $e^{-S_A|_\vartheta}$ predicted by the holographic model and the purity $e^{-S_A}$ given by the entanglement data. A straightforward option would be the mean squared error (MSE) loss
\eq{\label{eq:L_MSE}\scL_{\vartheta,\text{MSE}}=\avg_A(e^{-S_A|_\vartheta}- e^{-S_A})^2.}
The model prediction $e^{-S_A|_\vartheta}$ should be evaluated according to \eqnref{eq:EF6}, which involves the ensemble expectation $\E_{[\omega]\sim P_\vartheta}$. In practice, the expectation value can only be estimated by sampling a finite number of Weyl field configurations from the generative model $P_\vartheta$ and take the average
\eq{e^{-S_A|_\vartheta}=\frac{1}{N_\omega}\sum_{[\omega]\sim P_\vartheta}\langle{\textstyle\prod}_{\partial A}\phi\rangle_\omega,}
where $N_\omega$ denotes the number of Weyl field samples. With the help of modern GPU, $\langle{\textstyle\prod}_{\partial A}\phi\rangle_\omega$ can be computed in parallel efficiently. The sample size $N_\omega$ is thus ultimately limited by the GPU memory. In our case, $N_\omega$ ranges from 512 to 2048.

For any finite sample size $N_\omega$, the finite average $e^{-S_A|_\vartheta}$ will have a finite variance, which bias the MSE loss
\eq{\scL_{\vartheta,\text{MSE}}=\avg_A\bigg((\E e^{-S_A|_\vartheta}- e^{-S_A})^2+\frac{\var e^{-S_A|_\vartheta}}{N_\omega}\bigg),}
causing the parameter to converge to a wrong saddle point. The bias can be corrected by assigning a larger weight to the prediction with a higher precision, i.e. 
\eq{\avg_A \frac{(e^{-S_A|_\vartheta}- e^{-S_A})^2}{\var e^{-S_A|_\vartheta}},}
which can also be argued from the maximum-likelihood estimation. The variance is generally proportional to the square of the purity $\var e^{-S_A|_\vartheta}\propto (e^{-S_A|_\vartheta})^2$, which leads to the mean squared relative error (MSRE) loss
\eqs{\label{eq:L_MSRE}\scL_{\vartheta,\text{MSRE}}&=\avg_A \Big(\frac{e^{-S_A|_\vartheta}- e^{-S_A}}{e^{-S_A|_\vartheta}}\Big)^2\\
&=\avg_{A}\big(1-e^{S_A|_\vartheta-S_A}\big)^2.}
We numerically test the loss function by generating some data using a model with known parameters, and train new models with different loss function on the generated data to see if the parameter converges to the known result. Our test shows that \eqnref{eq:L_MSRE} indeed converges better compare to \eqnref{eq:L_MSE}. Therefore, we will use the MSRE loss function to train the model, as mentioned in \eqnref{eq:loss}.
%\eqs{\mathop{\mathrm{Var}} e^{-S_A|_\vartheta}&=\frac{1}{N_\omega}\mathop{\mathrm{Var}}_{[\omega]}\langle{\textstyle\prod}_{\partial A}\phi\rangle_\omega\\&\leq \frac{1}{N_\omega^2}\sum_{[\omega]}\langle{\textstyle\prod}_{\partial A}\phi\rangle_\omega^2\\&\leq \Big(\frac{1}{N_\omega}\sum_{[\omega]}\langle{\textstyle\prod}_{\partial A}\phi\rangle_\omega\Big)^2=(e^{-S_A|_\vartheta})^2.}

\section{Numerical Results}

\subsection{Fitting Entanglement Data with Static and Fluctuating Geometry}

We apply the proposed machine learning approach to learn the entanglement feature of a Majorana fermion chain of 32 sites (16 unit cells) with a relatively large central charge $c=8$. The entanglement data is partitioned into the training set and the test set that does not overlap. Within the training/test set, the data can be further classified by the number of subregions of the entanglement region, including the single-region, double-region, and triple-region entanglement. To demonstrate the effect of introducing gravitational fluctuations, we will compare two holographic models: (i) the \emph{fluctuating} model, i.e.~the model $e^{-S_A|_\vartheta}=\E_{[\omega]\sim P_\vartheta}\langle\textstyle\prod_{\partial A}\phi\rangle_\omega$ proposed in \eqnref{eq:EF6} with fluctuating geometries, (ii) the \emph{static} model, i.e.~the model $e^{-S_A|_\vartheta}=\langle\textstyle\prod_{\partial A}\phi\rangle_{\omega\equiv 0}$ with a fixed static geometry. We train both models using the MSRE loss in \eqnref{eq:loss}. The algorithm is implemented in the TensorFlow\cite{MartinAbadi2015TLMLHS} framework using the ADAM\cite{Kingma2014AMSO} optimizer. Upon convergence, the MSRE loss is evaluated on the test set to characterize the performance of the model. The result is summarized in \tabref{tab:loss}

\begin{table}[h]
\begin{tabular}{cc|ccc}
  \multicolumn{2}{c|}{Model} & static & static & fluctuating \\
  \multicolumn{2}{c|}{Training set} & single & single+double & single+double\\
  \hline
  \multirow{3}{*}{ \parbox[t]{2mm}{\rotatebox[origin=c]{90}{Test set}}}
  & single & $8.7 \times 10^{-6}$ & $2.1 \times 10^{-2}$ & $1.5 \times 10^{-3}$ \\
  & double & $1.1 \times 10^{-1}$ & $3.9 \times 10^{-2}$ & $5.7 \times 10^{-3}$ \\
  & triple & $7.5 \times 10^{-1}$ & $6.0 \times 10^{-1}$ & $3.1 \times 10^{-1}$ \\
\end{tabular}
\caption{MSRE loss on test sets for different models and training sets. The model can be static geometry or fluctuating geometry. The training set can include on single-regions or both single- and double-regions. The test set can be either single-, double-, or triple-regions separately.}
\label{tab:loss}
\end{table}

\begin{figure}[h]
\begin{center}
\includegraphics[width=0.94\columnwidth]{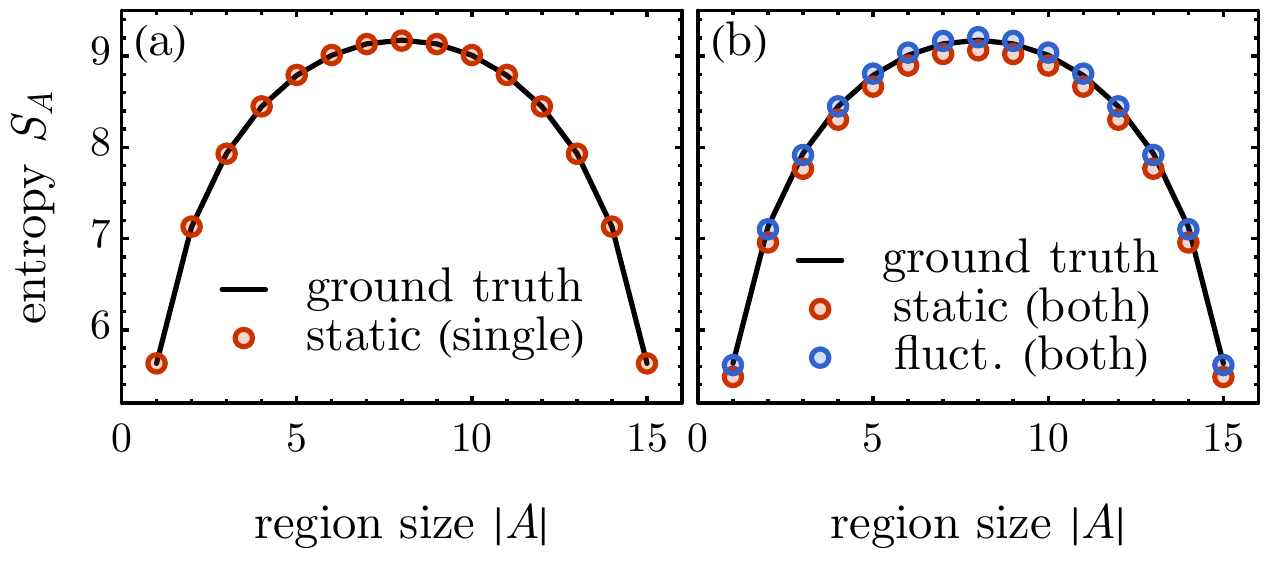}
\caption{Model predicted entanglement entropy v.s. ground truth for (a) static model trained on single-region entanglement data, (b) static and fluctuating models trained on both single-region and double-region entanglement data.}
\label{fig:entropy}
\end{center}
\end{figure}

If we train the static geometry model with single-region data only, the model can easily achieve high accuracy ($\text{MSRE}\sim10^{-5}$) in predicting single-region entanglement, as also shown in \figref{fig:entropy}(a). But the prediction of multi-region entanglement is rather inaccurate ($\text{MSRE}\sim10^{-1}$), meaning that the static geometry model overfits the single-region data and can not be generalized to multi-region data. If we include the double-region data in the training set, and train the static geometry model with both single- and double-region entanglement, the model will learn to predict double-region entanglement better at the price of losing the accuracy in predicting single-region entanglement, with the MSRE saturates at the $\sim 10^{-2}$ level. This implies an intrinsic conflict for the static geometry model in modeling the single- and multi-region entanglement simultaneously. 

However, by introducing gravitational fluctuations to the model, the fluctuating geometry model achieves one order of magnitude improvement in the prediction accuracy of both single- and double-region entanglement, as the MSRE drops to the $\sim 10^{-3}$ level, which is also manifest in \figref{fig:entropy}(b). This indicates that the gravitational fluctuation indeed helps to reconcile the conflict between single- and multi-region entanglements in the classical gravity model (RT formula) (see \appref{app:fluctuation} for an analytic analysis of how the conflict can be reconciled in principle). Moreover, the prediction accuracy in triple-regions is also improved significantly, even if the model is never trained on the triple-region data. This speaks for the better generalizability of the fluctuating geometry model.

\begin{figure}[h]
\begin{center}
\includegraphics[width=0.95\columnwidth]{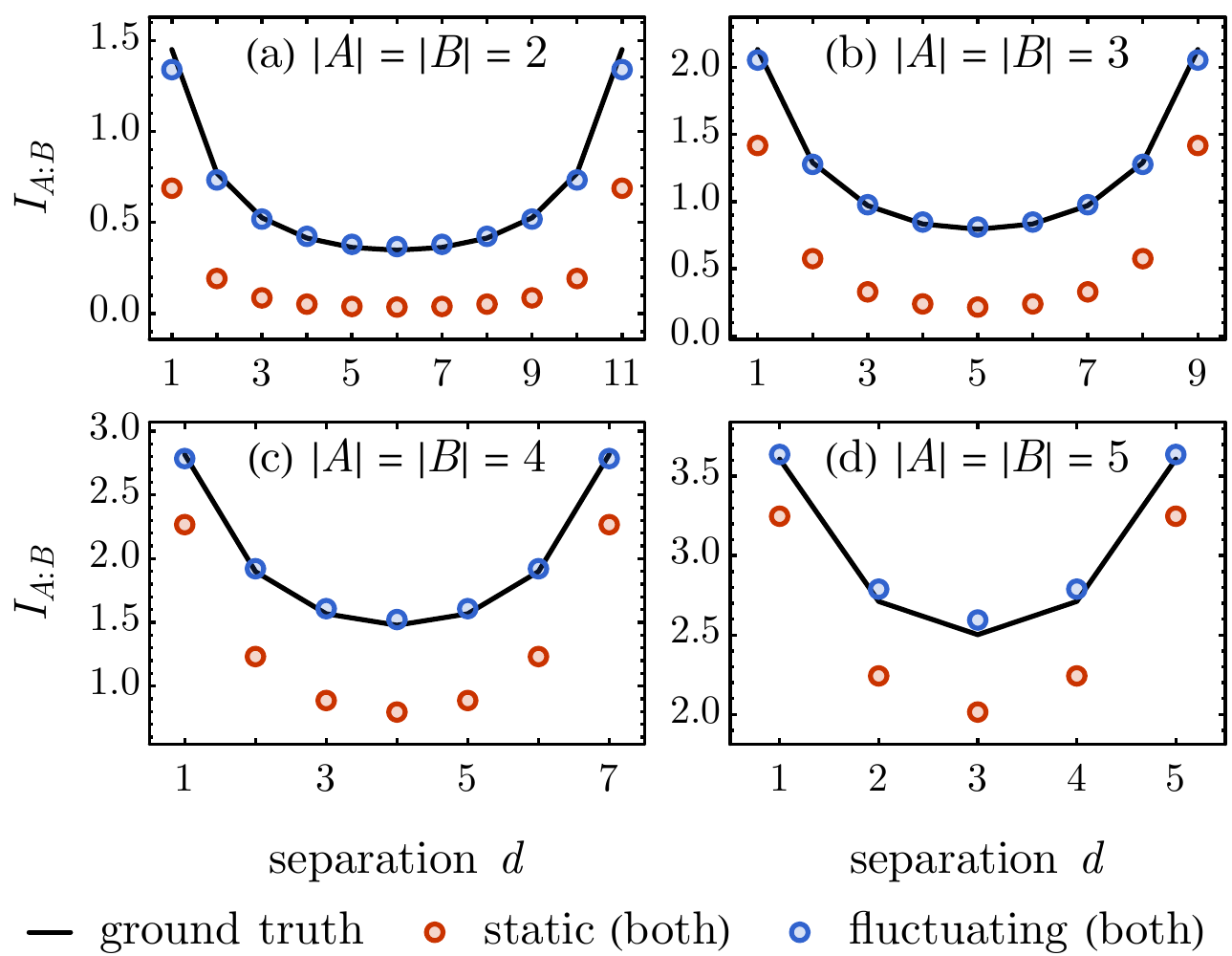}
\caption{Mutual information $I_{A:B}$ between two equal-sized regions $A$ and $B$ for different inter-region separations. Data points show predictions by the static model (red) and the fluctuating model (blue). Both models are trained with \emph{both} single- and double-region entanglement data.}
\label{fig:mutual_info}
\end{center}
\end{figure}

As argued previously, the static geometry model suffers from the problem of vanishing mutual information between far separated regions. One motivation to introduce gravitational fluctuations is to mediate the mutual information between distant regions through the holographic bulk. Indeed, as shown in \figref{fig:mutual_info}, by allowing the geometry to fluctuate, the model can better capture the behavior of mutual information. In particular, the static model fails to produce the non-vanishing mutual information between distant regions, which is most obviously seen in \figref{fig:mutual_info}(a), where the regions are most far separated (compare to their sizes). However, the fluctuating model fixes this problem, demonstrating the importance of introducing the geometric fluctuation in modeling the multi-region entanglement.

\subsection{Weyl Field Correlation and Effective Bulk Gravity Theory}

After training, we want to open up the model and see what bulk gravity theory has been learned. With the trained generative model $P_\vartheta[\omega]$ that describes the statistical fluctuation of the Weyl field $\omega$, we can explore various statistical properties of the distribution $P_\vartheta[\omega]$ to gain a deeper understanding of the optimal bulk gravity theory that emerges from learning the boundary entanglement data.

\begin{figure}[h]
\begin{center}
\includegraphics[width=0.96\columnwidth]{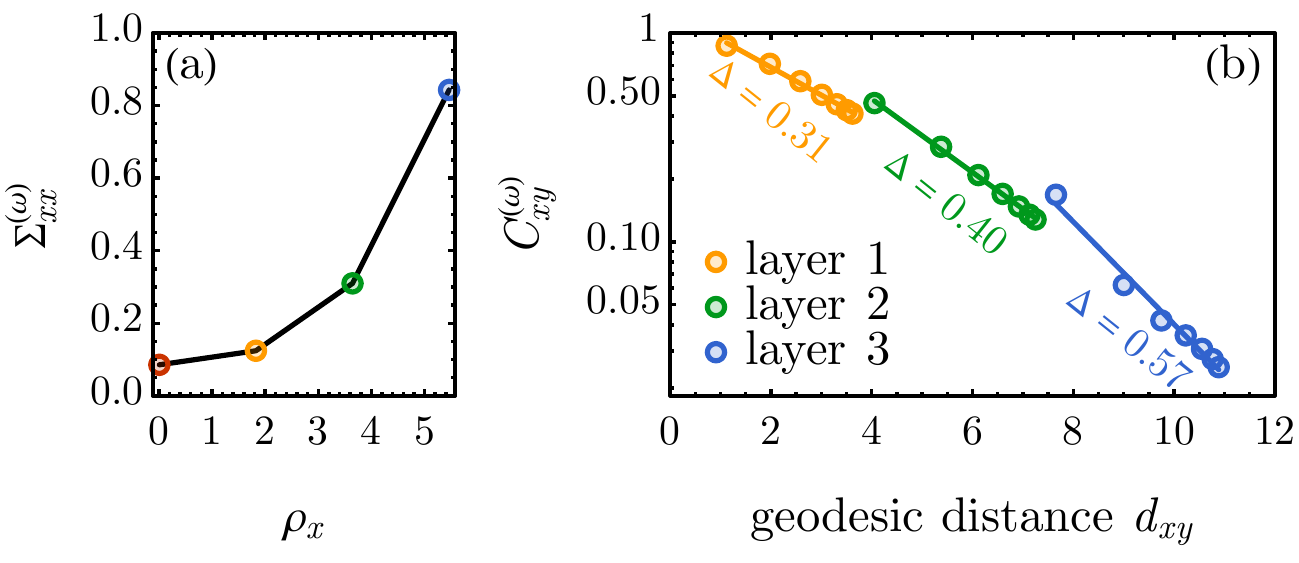}
\caption{(a) Weyl field local covariance $\Sigma^{(\omega)}_{xx}$ v.s. the radius coordinate $\rho_x$. (b) Intra-layer correlation $C^{(\omega)}_{xy}$ of the Weyl field (in logarithmic scale) v.s. the bulk geodesic distance $d_{xy}$, where $x$, $y$ points are taken from the same layer. The distance between points on a larger radius (or a higher layer) appears farther due to the hyperbolic background geometry, but the inverse correlation length $\Delta$ (the slope) remains roughly on the same order of magnitude across layers. }
\label{fig:Weyl}
\end{center}
\end{figure}

\begin{figure}[h]
\begin{center}
\includegraphics[width=\columnwidth]{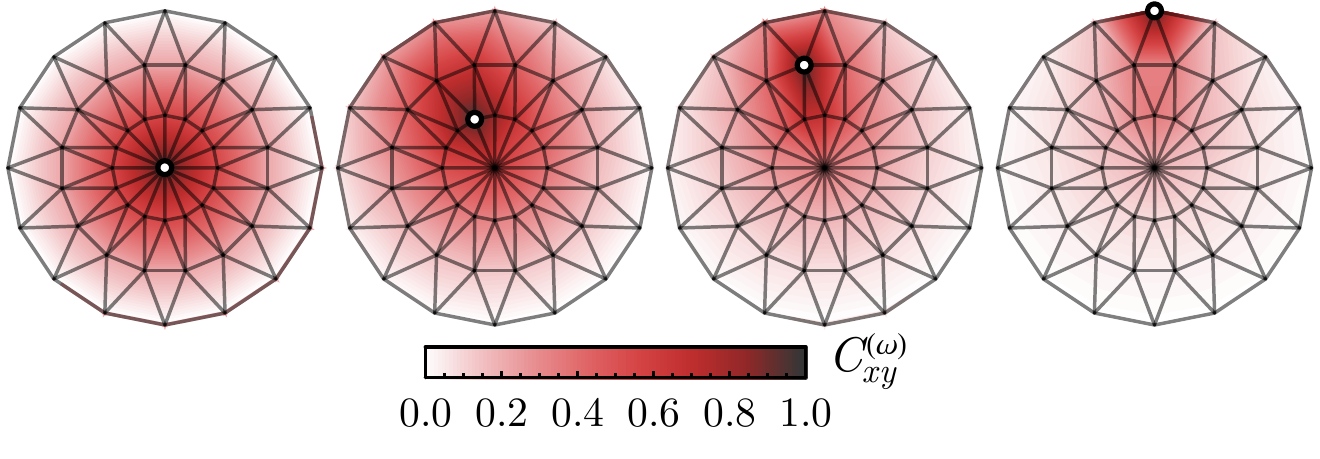}
\caption{Weyl field correlation in the bulk, between a marked reference point and the remaining points. Results are shown for the reference point placed in different layers.}
\label{fig:Weyl_correlation}
\end{center}
\end{figure}

\begin{figure}[h]
\begin{center}
\includegraphics[width=0.74\columnwidth]{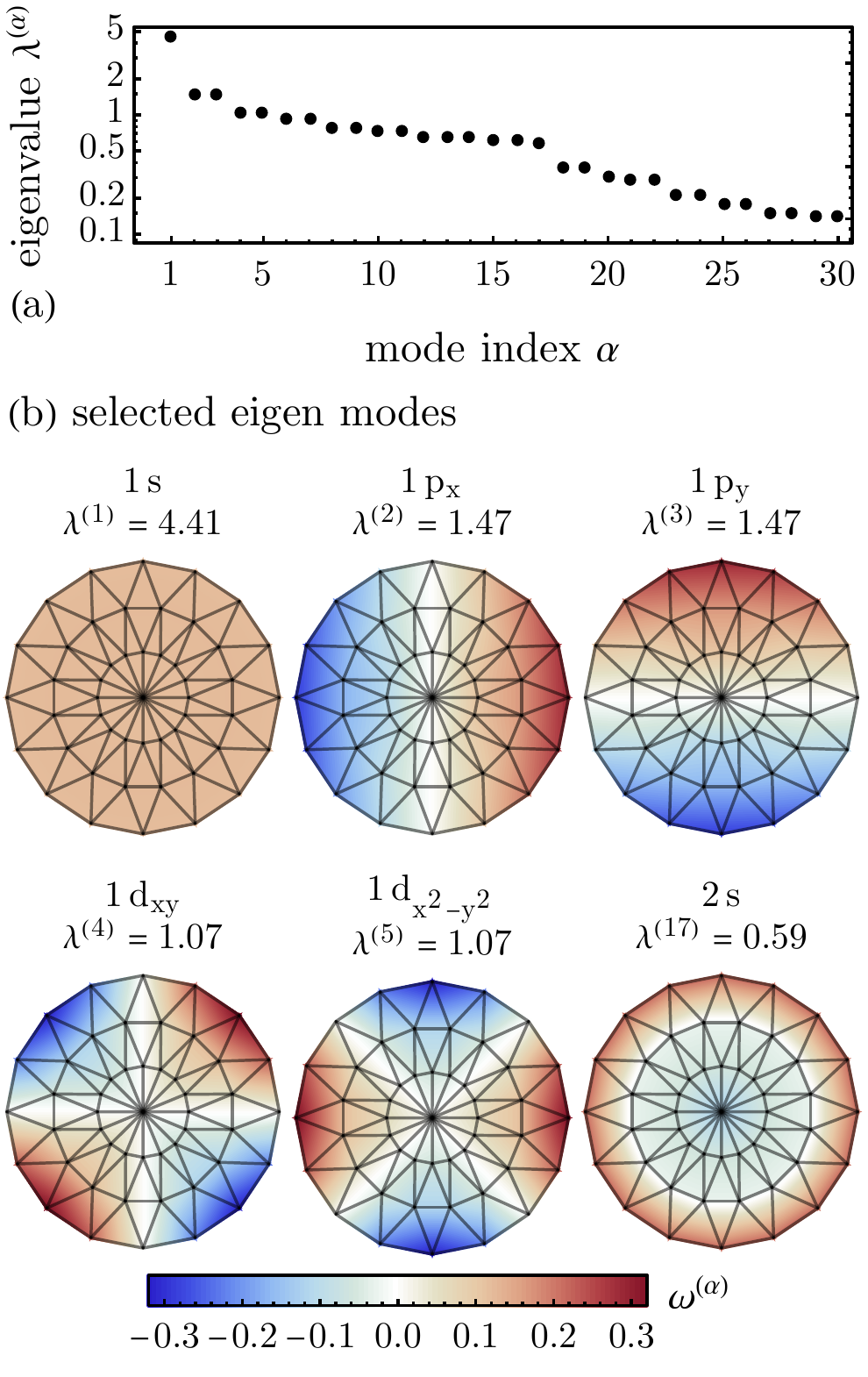}
\caption{Spectrum of the Weyl field fluctuation. (a) Leading eigenvalues $\lambda^{(\alpha)}$ of the covariance function (in logarithmic scale). (b) Selected eigenmodes $\omega^{(\alpha)}$ labeled by the principal and angular quantum numbers.}
\label{fig:Weyl_spectrum}
\end{center}
\end{figure}

We first study the covariance function $\Sigma^{(\omega)}_{xy}$ of the Weyl field $\omega$, defined as
\eq{\Sigma^{(\omega)}_{xy}=\E_{[\omega]\sim P_\vartheta}\omega_x\omega_y=\int_{[\omega]}P_\vartheta[\omega]\omega_x\omega_y.}
We observe that its diagonal elements $\Sigma^{(\omega)}_{xx}$ (i.e.~the local covariance) grows with the radius $\rho_x$ coordinate (as $x$ approaches the boundary), see \figref{fig:Weyl}(a). This is because the discretization scale is changing along the radius direction. In our discretization scheme as shown in \figref{fig:lattice}, the hyperbolic space is finer discretized towards the center of the bulk, therefore the field $\omega$ will appear to be stiffer near the bulk center, and hence its covariance is smaller. To eliminate this influence of the discretization scheme, we normalize (standardize) the covariance and define the correlation function
\eq{C^{(\omega)}_{xy}=\frac{\Sigma^{(\omega)}_{xy}}{\sqrt{\Sigma^{(\omega)}_{xx} \Sigma^{(\omega)}_{yy}}}.}
We found that the Weyl field correlation $C^{(\omega)}_{xy}$ decays exponentially with respect to the geodesic distance $d_{xy}$ in the holographic bulk
\eq{C^{(\omega)}_{xy}\sim \exp(-\Delta d_{xy}),}
where the inverse correlation length $\Delta$ remains almost the same across different layers in the bulk, as shown in \figref{fig:Weyl}(b). The short-ranged nature of the Weyl field correlation is more obviously shown in \figref{fig:Weyl_correlation}, which is an unequivocal sign of locality. In other words, the machine has learned from the entanglement data that the Weyl field fluctuation can be described by a local model (as the correlation is short-ranged) in the bulk. This emergent locality is remarkable since locality was never explicitly given to the generative model at the architecture level: the neural network in the generator $G_\vartheta$ was fully connected, which in principle allows non-local / long-ranged correlation of $\omega$ across the bulk, yet a short-ranged correlation emerges from learning the entanglement data.

Further more, we can learn about the leading modes of gravitational fluctuations in the machine-learned distribution $P_\vartheta[\omega]$ by computing the spectral decomposition of the covariance function
\eq{\Sigma^{(\omega)}_{xy}=\sum_{\alpha}\lambda^{(\alpha)} \omega^{(\alpha)}_x \omega^{(\alpha)}_y,}
where $\lambda^{(\alpha)}$ is the $\alpha$th eigenvalue and $\omega^{(\alpha)}$ is the corresponding eigenmode. The result is shown in \figref{fig:Weyl_spectrum}. The long wave-length collective fluctuations emerges as the leading (low-energy) modes of gravitational fluctuation automatically. Using the covariant function $\Sigma^{(\omega)}_{xy}$, one can reconstruct the effective gravitational action to the quadratic order (at Gaussian level)
\eq{S[\omega]=\frac{1}{2}\sum_{x,y}\omega_x \Sigma^{(\omega)}_{xy}\omega_y+\cdots,}
such that $P[\omega]\propto e^{-S[\omega]}$ approximately. In this way, the machine-learning model helps us to extract a statistical gravity theory (in terms of the Weyl field theory) from the entanglement data, demonstrating a data-driven approach to establish the holographic duality.

\subsection{Matter Field Mass Renormalization Effect}

\begin{figure}[h]
\begin{center}
\includegraphics[width=0.57\columnwidth]{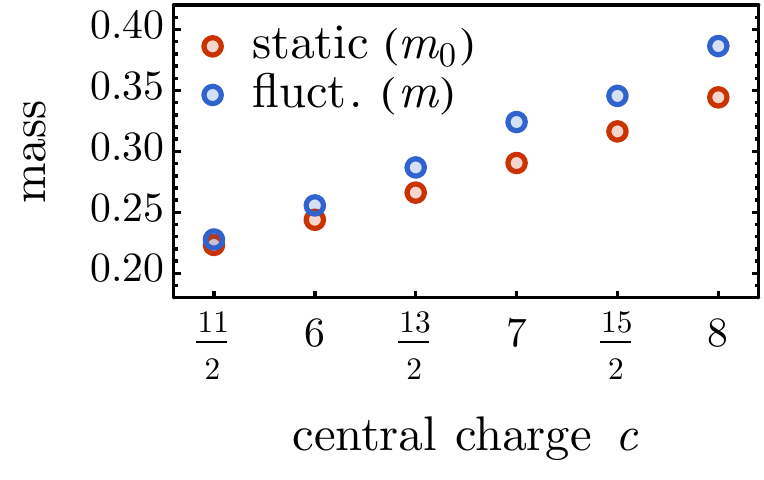}
\caption{Trained values of the scalar field mass $m$ for different central charge $c$ of the free fermion CFT, based on the static model (red) and the fluctuating model (blue).}
\label{fig:mass}
\end{center}
\end{figure}

As we have seen, geometric fluctuation effectively introduces interactions between the bulk scalar field $\phi$, which generates the desired behavior for mutual information. As a consequence, the bare mass $m$ of the scalar field should also be renormalized by the gravitational interaction. Remarkably, we can observe such a renormalization effect in our holographic model, by comparing the static model (without geometric fluctuation) and the fluctuating model (with geometric fluctuation). The mass parameter $m$ is trainable in both models, but their optimal values are different due to the renormalization effect. We train the static model on the single-region entanglement data, and the fluctuating model on both the single- and double-region entanglement data. For a range of total central charge $11/2\leq c\leq 8$ studied, we observe that the trained value of the (bare) mass $m$ in the fluctuating model is systematically larger than that $m_0$ in the static model, as shown in \figref{fig:mass}.

The mass renormalization effect can be understood heuristically by consider a single-region entanglement. In the static model, the entanglement entropy is modeled by $e^{-S_A}\sim e^{-m_0|\gamma_A|}$ where $\gamma_A$ is the geodesic connecting the entanglement cuts of $A$ through the static bulk. With geometric fluctuation $|\gamma_A|\to|\gamma_A|+\delta|\gamma_A|$ (where $\delta|\gamma_A|$ is the additional geodesic length due to the Weyl field $\omega$), the entanglement entropy will be modeled by
\eq{e^{-S_{A}}\sim \E_\omega e^{-m(|\gamma_A|+\delta|\gamma_A|)}\simeq e^{-m(|\gamma_A|-\frac{m}{2}\E_\omega(\delta|\gamma_A|)^2)}.}
For these two models to match, we must have $m>m_0$, which qualitatively explains our observation.

\section{Summary and Discussion}

We present a machine-learning approach to extract the holography statistical gravity theory from the data of multi-region entanglement entropy in a quantum many-body system. Our work advances both the field of tensor network holography and the field of machine learning holography. (i) On the tensor network holography side, we generalize the random tensor network (RTN) model to incorporate the bond dimension fluctuation, which makes the model more expressive in capturing features of multi-region entanglement. We derive the holographic bulk theory for the RTN with bond dimension fluctuation and show that the dual gravity theory consists of a massive scalar field on a fluctuating background geometry. The idea of using Ising duality in the derivation is also quite original, which provides an alternative view of the bulk theory that has not been presented in literature, as we are aware of. (ii) On the machine learning holography side, our work goes beyond the previous approaches\cite{Gan2017HDL,You2018Machine,Dong2018SOGNQSRQ,Hashimoto2018DLMC,Hashimoto2018DLH,Hashimoto2019ADBM,Hashimoto2020Neural} of inferring only a static background geometry from the boundary quantum data. By modeling the bulk geometric fluctuation with a generative model, our approach can extract a statistical gravity theory from the quantum entanglement data. Remarkably, we found that the machine-constructed gravity theory exhibit an emergent locality, which reveals the hidden bulk locality behind the non-local quantum entanglement on the boundary. 

Our work provides a novel data-driven approach to explore the emergent gravity from quantum entanglement. Combining with the recent development of efficient numerical methods to simulate entanglement dynamics in quantum many-body systems\cite{Kuo2019Markovian,Akhtar2020Multiregion,Fan2021Self-organized}, we can further explore the corresponding gravity dynamics in the holographic bulk, which will deepen our understanding of emergent gravity from quantum entanglement. On the practical side, our algorithm will boost the efficiency to model the entanglement structure of quantum many-body systems, which will find applications in quantum algorithm optimization and quantum circuit design.

\begin{acknowledgments}
We acknowledge the helpful discussions with John McGreevy, Xiao-Liang Qi, Zhenbin Yang. The authors are supported by a startup fund from UCSD and a UC Hellman Fellowship. 
\end{acknowledgments}

\bibliography{ref}% Produces the bibliography via BibTeX.
\newpage
\onecolumngrid

\appendix
\section{Perturbative Analysis of Mutual and Tripartite Information}
\label{app:perturbative}

The random tensor network model points to a bulk theory described by the following action
\eq{\label{eq:S_bulk_theory}S[\phi,\omega]=S[\phi|\omega]+S[\omega],}
where $S[\phi|\omega]=\frac{1}{2}\sum_{x,y}\phi_{x}K^{(\phi)}_{xy}[\omega]\phi_{y}$ follows from \eqnref{eq:S_phi|omega}, and we take a quadratic action $S[\omega]=\frac{1}{2}\sum_{x,y}\omega_{x}K^{(\omega)}_{xy}\omega_{y}$ for simplicity. In the perturbative limit, we assume that the fluctuation of the field $\omega$ is small, such that we can expand $K^{(\phi)}_{xy}[\omega]=K^{(\phi)}_{xy}+g\omega_x\delta_{x y}$, where $K^{(\phi)}_{xy}$ denotes the bare kernel of $\phi$ on the $\omega=0$ background. Therefore, the bulk theory becomes
\eq{S[\phi,\omega]=\frac{1}{2}\sum_{x,y}(\phi_{x}K^{(\phi)}_{xy}\phi_{y}+\omega_{x}K^{(\omega)}_{xy}\omega_{y})+\frac{g}{2}\sum_{x}\omega_x\phi_x^2.}
Define the field theory average as
\eq{\langle\cdots\rangle=\frac{1}{Z}\int_{[\phi,\omega]}\cdots e^{-S[\phi,\omega]},}
then the entanglement entropy $S_A$ of a signle-region $A$ that ends at the dual sites $(x_1,x_2)$ can be written as
\eq{e^{-S_A}=\langle\phi_{x_1}\phi_{x_2}\rangle.}
The entanglement entropy for multi-regions are modeled similarly as multi-point covariance of the $\phi$ field among all boundary points.

Now we consider three regions $A$, $B$ and $C$ boundaried by $(x_1,x_2)$, $(x_3,x_4)$ and $(x_5,x_6)$ respectively. The mutual information $I_{A:B}=S_A+S_B-S_{AB}$ and the tripartite information
$I_{A:B:C}=S_{A}+S_{B}+S_{C}-S_{AB}-S_{BC}-S_{AC}+S_{ABC}$ can be evaluated by the following ratios of covariance function
\eqs{\label{eq:I_def}e^{-I_{A:B}}&=\frac{\langle\phi_{x_1}\phi_{x_2}\rangle\langle\phi_{x_3}\phi_{x_4}\rangle}{\langle\phi_{x_1}\phi_{x_2}\phi_{x_3}\phi_{x_4}\rangle},\\
e^{-I_{A:B:C}}&=\frac{\langle\phi_{x_1}\phi_{x_2}\rangle\langle\phi_{x_3}\phi_{x_4}\rangle\langle\phi_{x_5}\phi_{x_6}\rangle\langle\phi_{x_1}\phi_{x_2}\phi_{x_3}\phi_{x_4}\phi_{x_5}\phi_{x_6}\rangle}{\langle\phi_{x_1}\phi_{x_2}\phi_{x_3}\phi_{x_4}\rangle\langle\phi_{x_1}\phi_{x_2}\phi_{x_3}\phi_{x_4}\rangle\langle\phi_{x_1}\phi_{x_2}\phi_{x_3}\phi_{x_4}\rangle}.}

\begin{figure}[htbp]
\begin{center}
\includegraphics[width=0.7\columnwidth]{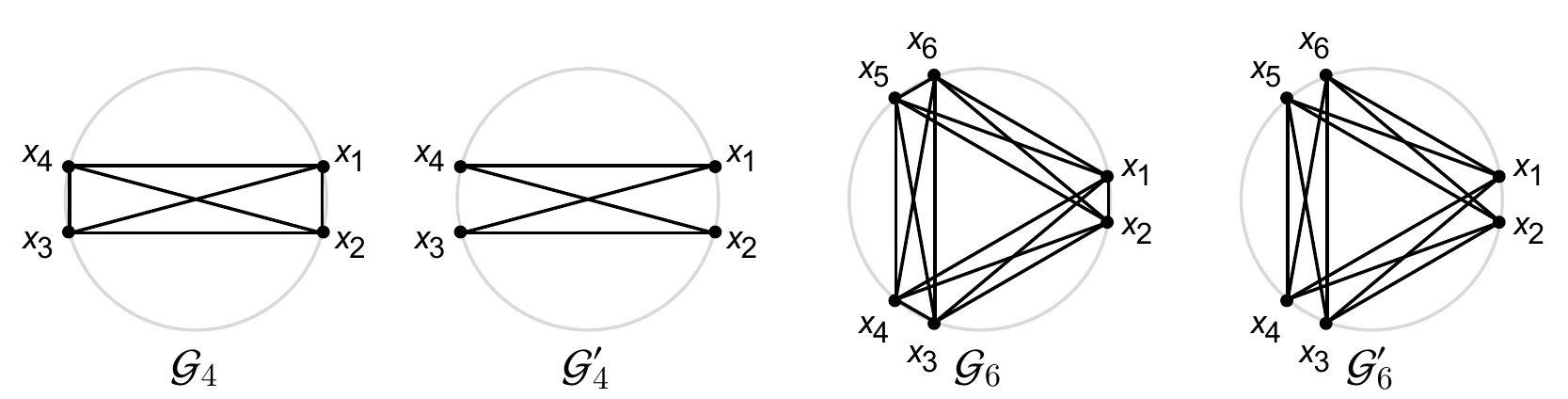}
\caption{A few useful graphs. }
\label{fig:graph}
\end{center}
\end{figure}

To simplify the notation in the following discussion, we introduce a few graphs in \figref{fig:graph}. Let $\scG_4$ be the complete graph over $x_1,x_2,x_3,x_4$, and $\scG_6$ be the complete graph over $x_1,x_2,x_3,x_4,x_5$. Further denote $\scG'_{2n}$ graph (with a prime) to be the graph with edges $(x_{2k-1}x_{2k})$ (for $k=1,\cdots, n$) removed from $\scG_{2n}$. Define the set of perfect matchings on a graph $\scG$ by $\scM[\scG]$ (where each perfect matching is a subset of edges such that every vertex is covered and only covered by one edge). We define the bare propagators (the covariance functions) $\Sigma^{(\phi)}$ and $\Sigma^{(\omega)}$ from the inverses of the bare kernels $K^{(\phi)}$ and $K^{(\omega)}$ for both $\phi$ and $\omega$ fields respectively,
\eq{\Sigma^{(\phi)}_{xy}=((K^{(\phi)})^{-1})_{xy}, \quad \Sigma^{(\omega)}_{xy}=((K^{(\omega)})^{-1})_{xy}.}
Using perturbative field theory (treating $g$ in \eqnref{eq:S_bulk_theory} as perturbation), to the 2nd order in $g$ (and keeping only the tree level diagrams), we can calculate the covariance functions
\eqs{\label{eq:corr2}\langle\phi_{x_1}\phi_{x_2}\rangle&=\Sigma^{(\phi)}_{x_1x_2},}
\eqs{\label{eq:corr4}\langle\phi_{x_1}\phi_{x_2}\phi_{x_3}\phi_{x_4}\rangle&=\sum_{(x_ix_j)(x_kx_l)\in\scM[\scG_4]}\bigg(\Sigma^{(\phi)}_{x_ix_j}\Sigma^{(\phi)}_{x_kx_l}+g^2\sum_{y_1,y_2}\Sigma^{(\omega)}_{y_1y_2}\Sigma^{(\phi)}_{x_iy_1}\Sigma^{(\phi)}_{y_1x_j}\Sigma^{(\phi)}_{x_ky_2}\Sigma^{(\phi)}_{y_2x_l}\bigg),}
\eqs{\label{eq:corr6}\langle\phi_{x_1}\phi_{x_2}\phi_{x_3}\phi_{x_4}\phi_{x_5}\phi_{x_6}\rangle&=\sum_{(x_ix_j)(x_kx_l)(x_mx_n)\in\scM[\scG_6]}\bigg(\Sigma^{(\phi)}_{x_ix_j}\Sigma^{(\phi)}_{x_kx_l}\Sigma^{(\phi)}_{x_mx_n}\\
&\hspace{120pt}+g^2\sum_{y_1,y_2}\Sigma^{(\omega)}_{y_1y_2}(\Sigma^{(\phi)}_{x_iy_1}\Sigma^{(\phi)}_{y_1x_j}\Sigma^{(\phi)}_{x_ky_2}\Sigma^{(\phi)}_{y_2x_l}\Sigma^{(\phi)}_{x_mx_n}\\&\hspace{180pt}+\Sigma^{(\phi)}_{x_iy_1}\Sigma^{(\phi)}_{y_1x_j}\Sigma^{(\phi)}_{x_kx_l}\Sigma^{(\phi)}_{x_my_2}\Sigma^{(\phi)}_{y_2x_n}\\&\hspace{180pt}+\Sigma^{(\phi)}_{x_ix_j}\Sigma^{(\phi)}_{x_ky_1}\Sigma^{(\phi)}_{y_1x_l}\Sigma^{(\phi)}_{x_my_2}\Sigma^{(\phi)}_{y_2x_n})\bigg).}

Substitute the correlation functions \eqnref{eq:corr2}-\eqnref{eq:corr6} to \eqnref{eq:I_def}, we find (to the $g^2$ order)
\eqs{\label{eq:IAB1}
I_{A:B}=&\sum_{(x_ix_j)(x_kx_l)\in\scM[\scG'_4]}\frac{\Sigma^{(\phi)}_{x_ix_j}\Sigma^{(\phi)}_{x_kx_l}}{\Sigma^{(\phi)}_{x_1x_2}\Sigma^{(\phi)}_{x_3x_4}}\\
&+g^2\sum_{(x_ix_j)(x_kx_l)\in\scM[\scG_4]}\sum_{y_1,y_2}\Sigma^{(\omega)}_{y_1y_2}\frac{\Sigma^{(\phi)}_{x_iy_1}\Sigma^{(\phi)}_{y_1x_j}\Sigma^{(\phi)}_{x_ky_2}\Sigma^{(\phi)}_{y_2x_l}}{\Sigma^{(\phi)}_{x_1x_2}\Sigma^{(\phi)}_{x_3x_4}},}
\eqs{
I_{A:B:C}=&-\sum_{(x_ix_j)(x_kx_l)(x_mx_n)\in\scM[\scG'_6]}\frac{\Sigma^{(\phi)}_{x_ix_j}\Sigma^{(\phi)}_{x_kx_l}\Sigma^{(\phi)}_{x_mx_n}}{\Sigma^{(\phi)}_{x_1x_2}\Sigma^{(\phi)}_{x_3x_4}\Sigma^{(\phi)}_{x_5x_6}}\\
&-g^2\sum_{(x_ix_j)(x_kx_l)(x_mx_n)\in\scM[\scG'_6]}\sum_{y_1,y_2}\Sigma^{(\omega)}_{y_1y_2}\Big(\frac{\Sigma^{(\phi)}_{x_iy_1}\Sigma^{(\phi)}_{y_1x_j}\Sigma^{(\phi)}_{x_ky_2}\Sigma^{(\phi)}_{y_2x_l}\Sigma^{(\phi)}_{x_mx_n}}{\Sigma^{(\phi)}_{x_1x_2}\Sigma^{(\phi)}_{x_3x_4}\Sigma^{(\phi)}_{x_5x_6}}+\frac{\Sigma^{(\phi)}_{x_iy_1}\Sigma^{(\phi)}_{y_1x_k}\Sigma^{(\phi)}_{x_jy_2}\Sigma^{(\phi)}_{y_2x_l}\Sigma^{(\phi)}_{x_mx_n}}{2\Sigma^{(\phi)}_{x_1x_2}\Sigma^{(\phi)}_{x_3x_4}\Sigma^{(\phi)}_{x_5x_6}}\\
&\hspace{170pt}+\frac{\Sigma^{(\phi)}_{x_iy_1}\Sigma^{(\phi)}_{y_1x_j}\Sigma^{(\phi)}_{x_kx_l}\Sigma^{(\phi)}_{x_my_2}\Sigma^{(\phi)}_{y_2x_n}}{\Sigma^{(\phi)}_{x_1x_2}\Sigma^{(\phi)}_{x_3x_4}\Sigma^{(\phi)}_{x_5x_6}}+\frac{\Sigma^{(\phi)}_{x_iy_1}\Sigma^{(\phi)}_{y_1x_m}\Sigma^{(\phi)}_{x_kx_l}\Sigma^{(\phi)}_{x_jy_2}\Sigma^{(\phi)}_{y_2x_n}}{2\Sigma^{(\phi)}_{x_1x_2}\Sigma^{(\phi)}_{x_3x_4}\Sigma^{(\phi)}_{x_5x_6}}\\
&\hspace{170pt}+\frac{\Sigma^{(\phi)}_{x_ix_j}\Sigma^{(\phi)}_{x_ky_1}\Sigma^{(\phi)}_{y_1x_l}\Sigma^{(\phi)}_{x_my_2}\Sigma^{(\phi)}_{y_2x_n}}{\Sigma^{(\phi)}_{x_1x_2}\Sigma^{(\phi)}_{x_3x_4}\Sigma^{(\phi)}_{x_5x_6}}+\frac{\Sigma^{(\phi)}_{x_ix_j}\Sigma^{(\phi)}_{x_ky_1}\Sigma^{(\phi)}_{y_1x_m}\Sigma^{(\phi)}_{x_ly_2}\Sigma^{(\phi)}_{y_2x_n}}{2\Sigma^{(\phi)}_{x_1x_2}\Sigma^{(\phi)}_{x_3x_4}\Sigma^{(\phi)}_{x_5x_6}}\Big).}
In the case that $\Sigma^{(\phi)}_{xy}\geq 0$ and $\Sigma^{(\omega)}_{xy}\geq 0$ (which is typically the case), we can ensure $I_{A:B}\geq 0$ and $I_{A:B:C}\leq 0$. The result proves that random tensor network model can produce a negative tripartite information $I_{A:B:C}$, which is a unique feature of quantum many-body entanglement that can not be achieved in classical systems. A negative tripartite information is an indication of quantum information scrambling and chaotic quantum dynamics in the quantum system. Although the bulk theory is a classical statistical gravity model, it can still model the quantum chaotic entanglement features on the holographic boundary, which speak for the strong expression power of the random tensor network model.

\section{Necessity and Expected Behavior of Weyl Field Fluctuation}
\label{app:fluctuation}

We would like to take a closer look at the mutual information. It would more intuitive to present the diagrammatic representation of \eqnref{eq:IAB1} 
\eq{\label{eq:IAB_diag}I_{A:B}=\raisebox{-34pt}{\includegraphics[height=72pt]{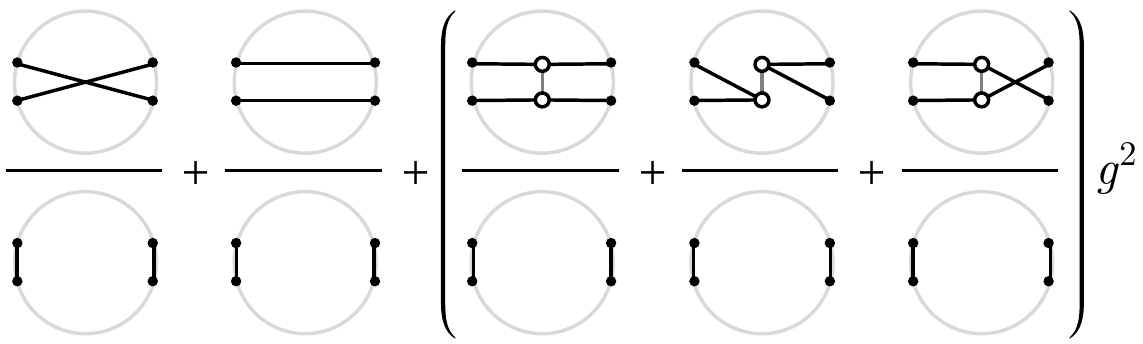}}.}
where points on the boundary correspond to $x_1,x_2,x_3,x_4$ (following the arrangement of vertices in the $\scG_4$ graph shown in \figref{fig:graph}) and the small circles in the bulk correspond to $y_1,y_2$ that should be summed over. The black lines represent $\Sigma^{(\phi)}_{xy}$ and the gray lines represent $\Sigma^{(\omega)}_{xy}$. The perturbation $g$ parameterizes the coupling strength of the bulk scalar field $\phi$ to the background gravitational fluctuation (the Weyl field $\omega$). Setting $g=0$ will decouple the gravitational fluctuation, which effectively corresponds to a static bulk model (because gravitational fluctuation will have no effect in the decoupled limit). Let us consider the case when regions $A$ and $B$ are far separated, meaning that the spacings  $|x_1-x_2|$ and $|x_3-x_4|$ are small. In this case, we expect the mutual information to decay with the inter-region spacing in a power-law manner with the power set by the smallest scaling dimension of the critical field in the quantum system on the holographic boundary, because the mutual information upper bounds all correlation function between regions $A$ and $B$, which can not decay faster than the lightest critical field. As we will see, this behavior can only be reproduce via the bulk model gravitational fluctuation is included.

To argue the necessity of including gravitational fluctuation, we first consider the decoupled limit (i.e.~$g=0$) to demonstrate why it fails to capture the correct behavior of mutual information. In the $g=0$ limit, the first two terms can still contribute to a finite mutual information that decays with the inter-region separation in a power-law manner, but the power will be set by the central charge of the quantum system on the holographic boundary. Because the power-law comes from the $\phi$-field correlation $\Sigma^{(\phi)}_{xy}$, whose scaling is determined by the single-region entanglement entropy, as $-\ln \Sigma^{(\phi)}_{x_1x_2}\sim S_A\sim c\ln |A|$ (such that $x_1,x_2$ are boundary points of the region $A$). However, the total central charge $c=N/2$ can be as large as we wish in the large $N$ limit. Therefore, although the first two terms (the $g^0$ terms) in \eqnref{eq:IAB_diag} can produce a power-law decay mutual information, but the power will typically be too large (i.e. the mutual information will decay too fast). This reflects the internal inconsistency in describing both single- and double-region entanglements using a holographic bulk model without gravitational fluctuation.

An obvious solution is to introduce a different field from $\phi$ to mediate the mutual information across the holographic bulk. Then we will have an independent freedom, such that we can tune its scaling dimension to match that of the lightest critical field. This is one major motivation to introduce the gravitational fluctuation (or to couple the scaler field $\phi$ to a fluctuating Weyl field $\omega$). As we turn on the coupling $g$, the mutual information will be dominated by
\eq{I_{A:B}=g^2\sum_{y_1,y_2}\Sigma^{(\omega)}_{y_1y_2}\frac{\Sigma^{(\phi)}_{x_1y_1}\Sigma^{(\phi)}_{y_1x_2}\Sigma^{(\phi)}_{x_3y_2}\Sigma^{(\phi)}_{y_2x_4}}{\Sigma^{(\phi)}_{x_1x_2}\Sigma^{(\phi)}_{x_3x_4}},}
whose long range behavior scales with $\sim g^2\Sigma^{(\omega)}_{y_1y_2}$. The Weyl field $\omega$ has a different propagator $\Sigma^{(\omega)}_{xy}$, which can be independently tuned to make $-\ln \Sigma^{(\omega)}_{xy}\sim 2\Delta_\text{min} \ln |x-y|$  with $\Delta_\text{min}$ being the smallest scaling dimension in the quantum critical theory. Here $|x-y|$ denotes the distance between two boundary points $x$ and $y$ measured using the bound metric. Translate $|x-y|$ into the bulk distance $d_{xy}$ assuming the bulk has a hyperbolic background geometry, we have $d_{xy}\sim\ln |x-y|$, which implies $\Sigma^{(\omega)}_{xy}\sim e^{-2\Delta_\text{min}d_{xy}}$. This indicates that Weyl field must be heavy in the bulk to produce the exponential decay of its correlation function with the bulk distance. Indeed, such a massive Weyl field fluctuation does emerge in the machine-learnt bulk gravity theory.

\end{document}